\documentclass[fleqn,12pt]{wlscirep}
\usepackage[utf8]{inputenc}
\usepackage[T1]{fontenc}
\usepackage{lineno}
\usepackage{setspace}
\usepackage{hyperref}
\usepackage{geometry} 
\usepackage{tabularx} 
\usepackage{caption} 
\usepackage{booktabs} 
\usepackage{multirow}
\usepackage{footmisc}
\usepackage{tablefootnote}

\title{AI for bureaucratic productivity: Measuring the potential of AI to help automate 143 million UK government transactions}

\author[1,2,*]{Vincent J. Straub}
\author[1]{Youmna Hashem}
\author[1,*]{Jonathan Bright}
\author[1]{Satyam Bhagwanani}
\author[1,3]{Deborah Morgan}
\author[1]{John Francis}
\author[1]{Saba Esnaashari}
\author[1,4]{Helen Margetts}
\affil[1]{Public Policy Programme, Alan Turing Institute, London, NW1 2DB, UK}
\affil[2]{Leverhulme Centre for Demographic Science, Nuffield Department of Population Health, University of Oxford, Oxford, OX3 7LF, UK}
\affil[3]{Department of Computer Science, University of Bath, Bath, BA2 7AY, UK}
\affil[4]{Oxford Internet Institute, University of Oxford, Oxford, OX1 3JS, UK}

\affil[*]{vincent.straub@ndph.ox.ac.uk, jbright@turing.ac.uk}


\begin{document}

\begin{abstract}

There is currently considerable excitement within government about the potential of artificial intelligence to improve public service productivity through the automation of complex but repetitive bureaucratic tasks, freeing up the time of skilled staff. Here, we explore the size of this opportunity, by mapping out the scale of citizen-facing bureaucratic decision-making procedures within UK central government, and measuring their potential for AI-driven automation. We estimate that UK central government conducts approximately one billion citizen-facing transactions per year in the provision of around 400 services, of which approximately 143 million are complex repetitive transactions. We estimate that 84\% of these complex transactions are highly automatable, representing a huge potential opportunity: saving even an average of just one minute per complex transaction would save the equivalent of approximately 1,200 person-years of work every year. We also develop a model to estimate the volume of transactions a government service undertakes, providing a way for government to avoid conducting time consuming transaction volume measurements. Finally, we find that there is high turnover in the types of services government provide, meaning that automation efforts should focus on general procedures rather than services themselves which are likely to evolve over time. Overall, our work presents a novel perspective on the structure and functioning of modern government, and how it might evolve in the age of artificial intelligence.  

\end{abstract}

\maketitle

\section*{Introduction}

Government services are at the heart of modern states. These services, such as registering a birth, applying for a pension, or updating a driving license, are central to the way rules and regulations are enforced, licenses for activities are provided, and benefits are distributed. They are essential to ensuring citizen health and well-being \cite{helliwell_empirical_2018}, maintaining the social contract \cite{devarajan_broken_2018}, and responding to global challenges like climate breakdown \cite{biesbroek_public_2018}. Some researchers have regarded the provision of public services as the definition of government itself, describing modern states as experts in `social sorting' \cite{lyon_surveillance_2007} or `decision factories'\cite{bovens_street-level_2002, peeters_administrative_2023}. 

The provision of these services is sometimes complex, often involving a chain of administrative and bureaucratic decision-making that may require public servants to, for example, analyse application documents, confirm identities, make judgments on eligibility, amongst many other steps. Such long bureaucratic chains can lead to long waits for services \cite{boyne_sources_2003}; and the amount of time they cost can also be considerable, both for government \cite{bright_generative_2024,florence_lives_2023,office_for_national_statistics_time_2024} and citizens \cite{lowrey_time_2021, executive_office_of_the_president_of_the_united_states_tackling_nodate}. Reducing the bureaucratic cost and overhead of service provision has been made a priority by many different governments \cite{e-estonia_e-governance_2020, cabinet_office_over_2022, us_government_accountability_office_government_2023}. 

Recently, the use of automated decision-making and artificial intelligence (AI) has been proposed as a way to make government more responsive, efficient, and fair \cite{margetts_rethink_2019, straub_artificial_2023, straub_multidomain_2023}, building on a long history of scholarly efforts to understand and improve public administration \cite{simon_administrative_1947}. Especially with the rise of generative AI as a product ready to be integrated into organisations, there is currently optimism around the idea that government services can be made faster and more citizen-oriented, both improving public satisfaction, and cutting costs and bureaucratic overhead \cite{bright_generative_2024}. 

However, while the general opportunity is evident, what is less clear is exactly where in government these new technologies can be applied, and hence where the potential savings are. Part of the reason for this is because, as yet, work in public administration has not fully adopted the empirical measurement of bureaucratic task structures in government as an object of study. The focus of this paper is to address part of this deficit, by describing the range of services central government offers, measuring the volume of transactions through these services, and estimating the extent to which these transactions are potentially automatable.  

Of course, this is not to say that government bureaucracy itself has been overlooked as an object of study. Max Weber's early writings on bureaucratic styles of government were one of the founding works in the field of political science \cite{constas_max_1958}, and continue to animate discussion today \cite{bouckaert_neo-weberian_2023}. More recently, work has sought to explain when and why bureaucrats engage in processes of policy change \cite{cohen_street-level_2021}, or to explore how policy implementation is delegated from political principals to bureaucratic agents \cite{cook_principal-agent_1989}, eventually ending at the `street level' of delivery \cite{hupe_street-level_2007}, to take just a few examples from a huge field. However, these works address the overall nature, structure, and functioning of government bureaucracy, rather than trying to measure the amount of bureaucratic processes undertaken, or to map their location. Alongside this work, the field of political economy has more directly tackled this question by exploring `size' of government. The amount of public spending is of course a core concern in this area \cite{kau_size_1981}, generating debate about the ideal size of the state \cite{meltzer_rational_1981}, the areas where it should operate \cite{becker_deadweight_2003}, and the extent to which that spending is `productive' \cite{office_for_national_statistics_public_nodate}: again to pick just a handful of examples from many. However, here, measurements have remained at a high level and tended to take a cost-based focus, addressing, for example, overall sizes of budgets or public spending in a given department or in the state overall. This work has not drilled down to the micro level of individual transactions. Finally, a variety of studies have addressed automation and digitisation in the context of specific bureaucratic processes, looking, for instance, at the extent to which easing administrative burdens improves take-up \cite{fox_administrative_2020} or turns citizens themselves into a type of government bureaucrat \cite{madsen_accidental_2022}. However, again, these studies have not sought to measure government as a whole, but rather typically focused on just one or two defined processes.    

\begin{table}[t!]
    \captionsetup{justification=raggedright, singlelinecheck=false, font=footnotesize}
    \renewcommand{\arraystretch}{1.4} 
    \begin{footnotesize}
    \caption{Glossary of government terms}
    \label{table:table1}
    \centering
    \begin{tabularx}{\textwidth}{lX}
        \toprule
        \textbf{Term} & \textbf{Definition} \\
        \midrule
        Government & The collective group of people and organisations which exercise executive power in a state. \\

        Organisation & A government organisation is
        any central government department, ministry, agency, or body responsible for the provision and delivery of public services. An individual organisation often manages a specific topic (e.g. national security) or sector of public administration. \\

        Service & A government or public service is an informational or transactional exchange between a government organisation and a service user that may need to be completed in a particular way in order to achieve a desired outcome (e.g., an application for a passport). \\

        Service user & An individual or a businesses applying for a public service. \\

        Transaction & An exchange of personal information, money, permission, goods, or services between a service user and an organisation that leaves a digital trace or record. A public service can involve single or multiple transactions.  \\

        Information-based service & A service that does not involve a decision of any sort, such as a guide that simply provides information to the user (e.g., the UK government service 'COVID-19: guidance and support'). Given that information-based services mainly serve a search and retrieve function, the majority of these types of services are already largely automated. \\

        Decision-based service & A service that involves service users exchanging information, money, permission, or goods with an organisation, requiring a bureaucratic decision to be made which usually results in a change to a government record. In our study, we conceptualise a decision-based service as being made up of a bundle of tasks that need to be completed by a service delivery professional. \\ 

        Task & An action-based activity involving decision-making undertaken by an individual or group of service delivery professionals to deliver a service. It may be routine or non-routine in nature and require analytic, cognitive, interactive or manual skills (see \cite{autor_skill_2003}). \\

        Service delivery professional & A government employee working in a particular organisation who is responsible for the delivery of services such as processing visas, passports, and driving licenses. \\
        \bottomrule
    \end{tabularx}
  \end{footnotesize}
\end{table}

Hence, despite considerable interest in both the way bureaucracy functions and its overall size, thus far research has not addressed our core concern in this paper, which is to measure the detailed, micro level of government bureaucratic decision-making, and to use this measurement as a way of thinking about the potential for AI and automation within government. This level of measurement, which tries to get to the heart of both the volume and nature of decisions being made, has been neglected, in our view, partly because of lack of access to data and partly due to lack of theoretical interest. Both of these factors are now changing: the increasing digitisation of government processes \cite{margetts_rethinking_2022} has meant that granular data on transactions is now available in good quantities; and the rise of AI (and other types of automation processes) has generated theoretical interest into what extent bureaucratic transactions could be either partially or fully automated by these emerging technologies \cite{cabinet_office_government_nodate}.

Here we seek to capitalise on this possibility by (1) providing a study of the structure, volume, and dynamics of UK government service transactions, and (2) offering a concrete measure of their potential for AI adoption, building on existing measures of such potential in the private sector. Specifically, we analyse citizen-facing public services ($n=377$) across ($n=57$) central UK government organisations (i.e., ministries, departments, and agencies). We first quantify the distribution of services across organisations, the average volume of service transactions, and describe inequalities in transaction dynamics. We then estimate the technical feasibility of partly automating services based on their task attributes, as well as the potential for generative AI to benefit low-automatable services. Overall the paper presents a picture of the high potential for automation of transactions across many government decision-based services. 

\section*{Quantifying UK government service delivery}

\begin{figure}[!htbp]
\includegraphics[width=\textwidth]{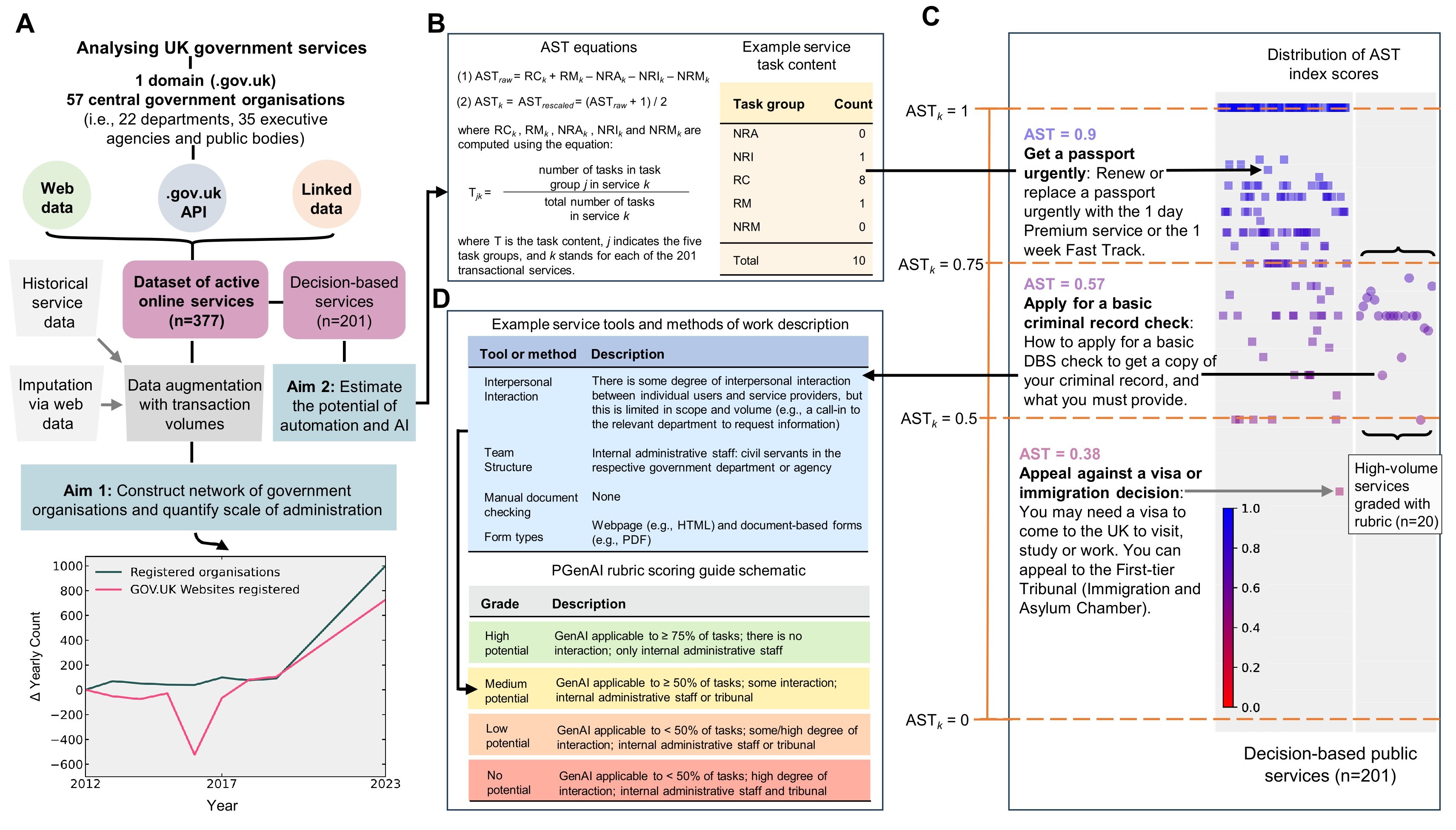}
\caption{\textbf{Overview of the data collection process and measurement approach.} This figure shows a schematic visualisation of the data collection process, including which data sources were accessed and how, alongside an explanation of the AST index and PGenAI rubric. (\textit{A}) The data collection workflow including the methods of data collection, preprocessing steps, and the two main study questions. The bottom panel plot shows the change in yearly count of number of government organisations registered on .gov.uk and change in total number of .gov.uk websites. (\textit{B}) The equation for the AST index used to measure automatability of decision-based services, and an example illustration of the task content for the service 'Get a passport urgently'. (\textit{C}) Distribution of AST index values for decision-based services ($n=201$) including high-volume services graded using PGenAI rubric ($n=20$), squares represent decision-based services and circles represent a subset of high-volume, low-automatable services, respectively (with axis values shown in orange on the left). AST index values of three UK government services included in our sample are highlighted for illustration, measured as of latest year of recorded service existence (2023). (\textit{D}) An example of the tools and methods of work description for the high-volume service `Apply for a basic criminal record check', used along with the service task content in assigning a PGenAI rubric score according to the scoring guide, depicted in schematic form in the bottom of the panel.}
\label{fig:fig1}
\end{figure}

The UK government is comprised of central departments, ministries, agencies, and executive bodies - herein referred to as organisations (\hyperref[table:table1]{Table~\ref*{table:table1}}) - that are designed to administer a diverse array of services to the public. As of the period between 2012 and 2024, the government encompassed 57 central departments, further supported by agencies dedicated to specific functions and objectives (this does not include local government bodies and services like the NHS). The workforce driving these operations consists of approximately 500,000 civil servants\cite{office_for_national_statistics_public_2023}. Notably, the majority of these civil servants (55.7\%) are engaged in service delivery \cite{cabinet_office_statistical_2023}, underscoring the importance of the government's role as a provider of public services. 

The primary units of analysis in our study are the transactions undertaken by central government departments and the organisations that work alongside them to deliver services. Within services, we delineate between information-based services (\hyperref [table:table1]{Table~\ref*{table:table1}}) and services which require administrative and bureaucratic decision-making. We define the latter services as decision-based services (\hyperref [table:table1]{Table~\ref*{table:table1}}). The volume and scope of service transactions, and the extent to which decision-based services present potential for automation, are our key focus areas in this study. 

Our data collection workflow is represented in Figure \ref{fig:fig1}, Panel A. Data on the size and scope of service transactions was assembled from four publicly-available sources: the main UK government website \href{.gov.uk}{.gov.uk}, a UK government open data repository called \href{data.gov.uk}{data.gov.uk} \footnote{data.gov.uk (\href{https://www.data.gov.uk/}{https://www.data.gov.uk/}) is the UK government's open data portal, available since January 2010. .gov.uk (\href{https://www.gov.uk/}{https://www.gov.uk/}) is the central point of digital access to government services, launched in January 2012.}, the Office for National Statistics (ONS), and the Government Digital Service (GDS), a unit of the Cabinet Office tasked with transforming the provision of online public services. The initial data collection phase involved collecting data from \href{.gov.uk}{.gov.uk} and \href{data.gov.uk}{data.gov.uk}, and requesting administrative data from the GDS. Via link sharing and direct download, we collected several datasets on the performance of historical (i.e., retired) government services and related webpages covering the period 2012-2023, including data on the number of .gov.uk domain names and open government websites (see bottom panel in \hyperref[fig:fig1]{Fig.~\ref*{fig:fig1}\textit{A}), as well as data on transaction volumes for $n=377$ active services (and $n=457$ historical services, Fig. \textit{S3-S4}}) and unique visits to $n=43,215$ service webpages, which in 2022 received a combined total of $n=1,631,836,608$ visits. 

Due to missing records on transaction volumes for the period after 2017, reflecting changes in the remit of GDS and the difficulty of standardising open data within government \cite{boyd_stuff_2017}, the second phase of data collection involved creating and collecting data on an analytical sample of government services (i.e., available online at the time of data collection). As there is no official publicly-available record of services, we first scraped all names of the $n=377$ active services and their respective .gov.uk URLs listed on a community-maintained list created by civil servants working on digital services\footnote{This site was recommended to us by GDS and is available at: \href{https://govuk-digital-services.herokuapp.com/}{https://govuk-digital-services.herokuapp.com/}.}. Although not exhaustive, it nevertheless relies on what we believe is the best possible current and publicly available record of services offered by the UK government at the time of writing. We then collected additional data by scraping and parsing the .gov.uk webpage for each service. Finally, we accessed the UK government Search API\footnote{Available at: \href{https://www.api.gov.uk/}{https://www.api.gov.uk/}.} to collect further descriptive data about services. After linking, cleaning and subsetting all datasets, our final analytical sample consisted of a list of $n=377$ services with an online web presence on .gov.uk. These services are the focus of the present study.

\subsection*{Volume and distribution of service transactions}

A second key task of the study was to estimate the volume of service transactions that take place each year. The workflow for achieving this is also set out in Figure \ref{fig:fig1}, Panel A. 

We rely on several sources of proxy data for our measurements of transaction volumes. First, the government made available a `Performance Platform' up until 2017 that contained data on volumes of transactions passing through some central government services that we are studying here\footnote{A copy of the platform was still available here at the time of writing: \href{https://webarchive.nationalarchives.gov.uk/ukgwa/20210315084942/https://www.gov.uk/performance/services}{https://webarchive.nationalarchives.gov.uk/ukgwa/20210315084942/https://www.gov.uk/performance/services}}. We matched services that were represented on this platform to services in our list of 377\footnote{Sometimes these figures were presented at a quarterly level, in which case we multiplied them by four, which is a potential source of inaccuracy as some services would be expected to have transaction volumes that fluctuate throughout the year.}. Second, the team behind this platform also shared with us a dataset of historical performance statistics for the period 2012-2017\footnote{We thank in particular Clifford Sheppard for sharing the data with us.}. In cases where a service in our list did not match to the 2017 version of the platform, we sought to match it to the most recent available figure from this dataset (which is available with the data release for this study). Finally, for services that were still unmatched, a web search was conducted to see if statistics were available for transaction volume (though these were only available in a few cases). In all of these cases, we matched services by their name first of all, and if they did not match we looked at the history of the service to see if it reflected the modification of a previously existing service. This matching was performed by one team member and then validated by two other team members. The data released alongside this study specifies where each individual transaction volume number was sourced from. 

This process resulted in 118 of the 377 services being assigned a transaction volume (though most of the highest volume services have a match, as we will show below). This relatively low number of matches is reflective of the high volumes of turnover within central government services. At the time of study, many services had appeared that didn't exist just a few years before, relating to (for example) the Covid-19 pandemic or the UK's departure from the European Union. Other services had been substantially changed or restructured, for example the UK's welfare system recently made a transition to `Universal Credit'. It is worth noting that this high turnover in services available is one potential barrier to easy automation: we will return to this point below. 

In order to assign transaction volume numbers to those services that were not matched by the above method, we developed a multilevel linear regression model to estimate transaction volumes based on the volume of visits the given service webpage received (as described in the previous section), and the type of service being provided. The model has a reasonably good fit (with an $R^2$ of 0.81), though of course these additional numbers should be treated only as rough estimates, a point that we will also pick up below. The fact that service transaction volume can be approximately estimated from web visit volumes is noteworthy, as it suggests that government may not need to invest a lot of time in collecting detailed transaction data (which has proved difficult to collect, as shown by the discontinuation of the Performance Platform after 2017), but can simply use web visits as a proxy.

\subsection*{Automatability of service transactions}

In addition to measuring the scope and volume of government service delivery, a further aim of the paper is to assess the automation potential of services, herein referred to as their automatability. Here, we specifically focus on assessing the automatability of decision-based services. The focus on this sub-section of services, as opposed to all service types, is based on the characteristics of the transactions required to deliver decision-based services. Decision-based service transactions absorb time and effort within government departments as they tend to require extensive chains of bureaucratic, often repetitive, procedures. As will be described in further detail below, these characteristics make them more amenable to measures of automation. After manually reviewing all 377 services, 201 were identified as decision-based services\footnote{The criteria used to assess whether a service was decision- or information-based follows the definitions provided in (\hyperref[table:table1]{Table~\ref*{table:table1}}).}. The remainder were classified as either information-based (and therefore already fully-automated), not having enough available data to inform a classification, or requiring a bespoke government account to access. 

To quantify the automation potential of decision-based services, we draw inspiration from the field of labour economics, which has developed a range of methodologies for the measurement of the potential for automation that new technologies present (though to our knowledge this has yet to be applied to government in detail). Developed to assess the impact of technological change on occupation polarisation, these approaches have expanded in recent years to inspire a body of research looking to measure and estimate the impact of computerisation and automation on labour markets. In this paper, we make use of the routine-biased technological change (RBTC) model introduced by Autor et al. \cite{autor_skill_2003}. This model proposes that, at a high level, different tasks required to perform occupations can be classified as either `routine' or `non-routine', depending on the extent to which they are repetitive and rules-based, the complexity of the task involved, and the extent to which they are creative or not. As routine tasks are easier to express in `explicit programmed rules' \cite{autor_skill_2003}, they are likely to be easier to computerise.  

According to this model, the share of routine tasks--or `routine task intensity' (RTI)-- within a given occupation indicates its potential for computerised automation. This focus on subdivision of occupations into tasks, with a subsequent measurement of the extent to which these tasks have a potential to be automated, has been the foundation upon which more recent studies of the likely impact of both machine learning \cite{brynjolfsson_what_2017, frey_future_2017}, and later generative AI \cite{frank_toward_2019} on the work force have been built. However, the majority of this work has taken the occupation as the main unit of analysis, with the occupation being a bundle of tasks that are more or less amenable to automation. In addition to applying this work to government, our work is distinctive by focusing on service transactions as bundles of tasks, rather than the people who help carry out the transactions. This focus on tasks allows us to explore the potential of applying AI and machine learning for a specific aspect of service delivery \cite{bozeman_public_2004}, without assuming this requires entire occupation roles to be automated away. Decision-based services such as the ones we study lend themselves particularly well to the task-based approach as they are made up of action-based activities involving micro decision-making that are more easily defined and measured than information-based services \cite{walker_non-transactional_2017}.  

To measure the automatability of decision-based services, we manually assign task statements (descriptions of specific tasks) to the 201 identified decision-based services. Task statements are sourced from the list of 193 tasks contained within the International Standard Classification of Occupations (ISCO)\cite{international_labour_organisation_isco-08_nodate}. Drawing on prior economics literature, the tasks are classified according to their task content \cite{fernandez-macias_comprehensive_2022}, and to their share of routine and non-routine content \cite{mihaylov_measuring_2019}, giving us five meta-task categories in total: non-routine analytic, non-routine interactive, routine cognitive, routine manual and non-routine manual tasks. 

In order to assign tasks to services, each service was manually reviewed by at least one of the authors. This review involved looking at the webpage associated with the service, reviewing any guidance and instructions provided about what it will take to complete the service, and following the workflow of completing the service. Tasks were assigned on the basis of what government itself would have to do to complete the service transaction. Here, the citizen-facing (or applicant-facing) tasks were used as proxies to identify what tasks would need to be completed in order to deliver the service.  

To calculate service automation potential we largely follow the procedure of Antonczyk and Fitzenberger \cite{antonczyk_can_2009} and Mihaylov and Tijdens \cite{mihaylov_measuring_2019}, who compute routine indexes by dividing the number of tasks in each task category by the total number of tasks across all categories (\hyperref[fig:fig1]{Fig.~\ref*{fig:fig1}\textit{B}}). This measure, which we label the Automatability of Service Tasks (AST) index, ranges between $-1$ and 1. To aid in interpretability, we rescale the AST distribution so that the range is $[0, 1]$, where a 1 can be interpreted as indicating that a service $k$ contains only routine tasks, and $0$ indicates that a service $k$ contains only non-routine tasks. \hyperref[fig:fig1]{Fig.~\ref*{fig:fig1}\textit{B-C}} present both the overall AST equation and also an example of the AST calculation for the service `Get a passport urgently'.  

\subsection*{Potential for Generative AI}
\label{sec:pgai}

In addition to estimating the potential automatibility of decision-based services, we also consider the extent to which the emergence of generative AI could potentially be used to enhance the delivery of the tasks we are reviewing. We use generative AI to refer to all types of systems built using AI techniques that can be used to create new text, images, video, audio, or code, and which can be expected to be capable of performing a variety of different government service tasks using little to no task-specific labeled training data. When accessed by civil servants, we expect generative AI to disrupt the task-specific paradigm of prior computer technologies used for automating government service tasks. While RBTC models capture a specific facet of technological change (computerised technology's ability to capture and automate structured, routine, and codifiable tasks) they may not capture the attributes and effects of this latest class of AI technology which might be capable of completing tasks previously thought to be `non-routine' \cite{frey_generative_2023}. Hence, to estimate the potential of generative AI for augmenting a service requiring subjective judgment or human interaction, we additionally propose a new rubric - the Potential for Generative AI (PGenAI) rubric - inspired by prior attempts to measure the workforce implications of supervised machine learning \cite{brynjolfsson_what_2017}. Specifically, we consider whether generative AI can support civil servants with the service delivery tasks associated with a particular service, including case management, responding to public requests, and maintaining public records. 

Our rubric, which grades services as having either `No potential', `Low potential', `Medium potential', or `High potential' (see \hyperref[fig:fig1]{Fig.~\ref*{fig:fig1}\textit{D}}), measures the overall exposure of services to generative AI based on the tasks that make up each service, following the spirit of prior work on quantifying potential suitability of supervised machine learning to occupations \cite{brynjolfsson_what_2017}, and more recently, large language models \cite{eloundou_gpts_2023}. We define potential as a proxy for likely impact without distinguishing between augmentation or automation effects, or what is termed labour-augmenting or labour-displacing in labour economics \cite{brynjolfsson_turing_2022}. Additionally, we follow the advice of more recent, comprehensive taxonomies of tasks for assessing the impact of new technologies on work \cite{fernandez-macias_comprehensive_2022} by also considering the methods and tools of work that facilitate the completition of these tasks, which here refer to: (1) the ways service delivery work is organised and (2) the physical objects used for aiding the production or service provision process (\hyperref[fig:fig1]{Fig.~\ref*{fig:fig1}\textit{D}}). We consider these additional two aspects as they are less dependent on what is being produced and more on the technology and social organisation of production. Therefore, they are more historically and institutionally contingent. 

\section*{Results}

In this section we will present the results from our analysis. We begin by exploring the breakdown of services offered, and the estimated transaction volumes going through them per year. 

\subsection*{Types and volumes of UK government transactions}

\begin{figure}[!htbp]
\centering
\includegraphics[width=.93\textwidth]{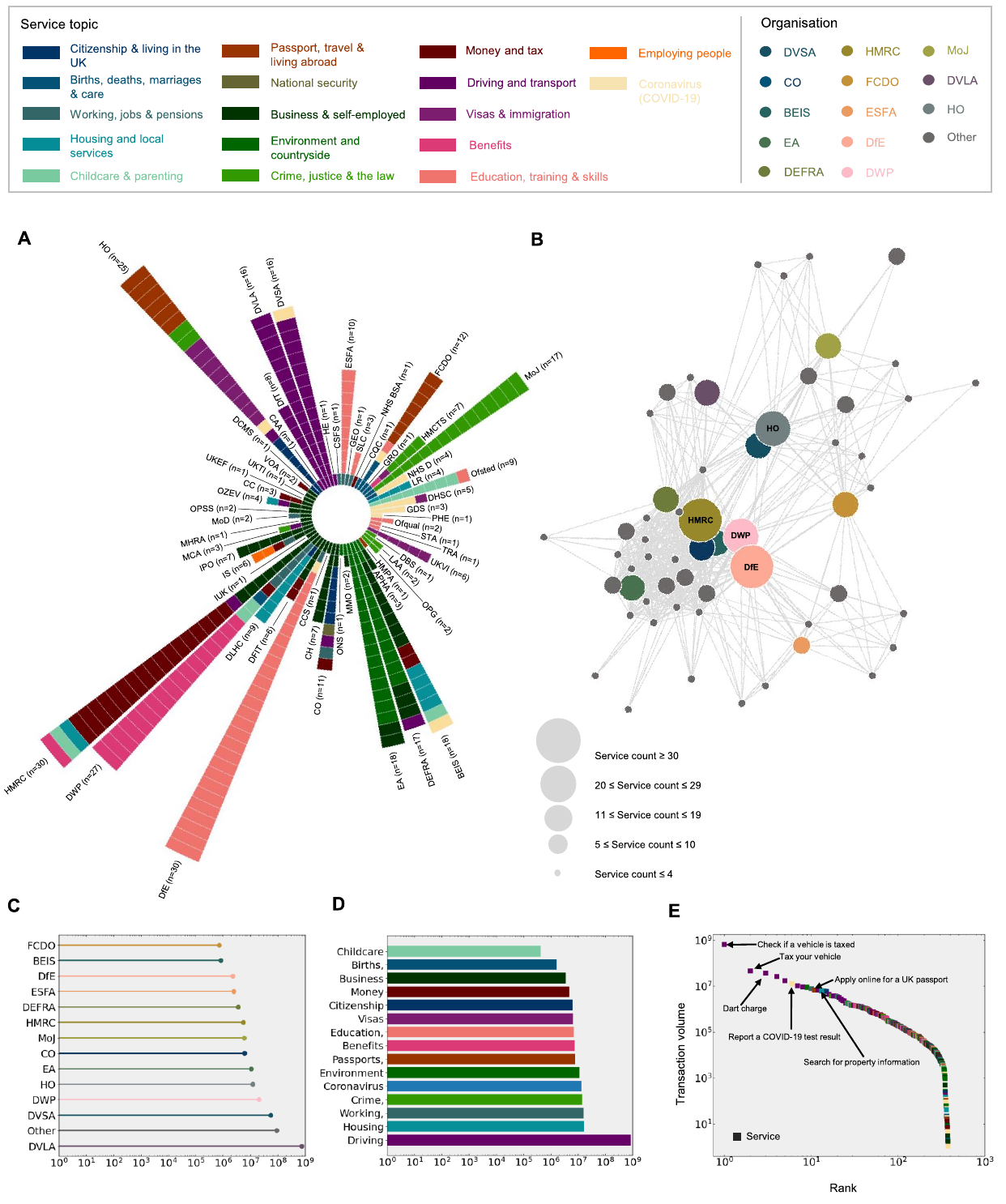}
\caption{\textbf{Quantifying the scale of UK government and public service transaction dynamics.} (\textit{A}) Distribution of service counts across all government organisations categorised by service topic (numbers in brackets indicate total service count). (\textit{B}) Structured network of UK government organisations. Node size indicates the count of services offered by a government organisation, arranged from center to periphery according to count of shared service topics between organisations. (\textit{C}) Distribution of transaction volumes for government organisations ($n=13$) with 10 or more services; remaining organisations are grouped under label `Other'. (\textit{D}) Distribution of transaction volumes across all service topics. (\textit{E}) Distribution of service transaction volumes ranked in descending order,  with selected services annotated. A legend is provided at the top of the figure for all panels.}
\label{fig:fig2}
\end{figure}

As described above, in our work we identify a total of 377 services offered by 57 UK central government organisations with at least some online component. This list is the main focus of analysis in the paper. We will begin by describing how these services are distributed. Figure \ref{fig:fig2}, Panel A shows how these services are spread across both government departments and official 'service topics' (n=17) that have been chosen and assigned to services by the UK government. HM Revenue and Customs (HMRC) and the Department for Education (DfE) stand out as departments with the highest volume of services (n=30), with the Home Office (HO) close behind (n=25). Amongst these service topics, `business and  self-employed' are the largest category (in terms of the number of services), whilst `driving and transport' and `education, training and skills' follow closely behind. However, the service topics list helps illustrate the wide range of services available across departments and organisations.

Figure \ref{fig:fig2}, Panel B shows that, despite this variety, there is also a lot of commonality between the topics of service being offered between certain departments. This figure looks at our service data in the form of a network, with each node representing a department, the size of each node being the number of services they offer (only departments with at least 15 services are represented), and the proximity of nodes representing the number of shared service topics. This way of looking at shared service topics provides one way of thinking about how to prioritise automation, as benefits from potential automation gains may be spread out across some departments more than others (for example, the most commonly overlapping service topic is 'business and self-employed': automating services in this area may provide disproportionate gains).  

Figure \ref{fig:fig2}, Panel C explores the distribution of service volumes in the dataset. In total we estimate that the 377 services we study undertake approximately 965 million transactions every year; however, this figure is highly skewed (see Figure \ref{fig:fig2}, Panel E), with one service ('Check if a vehicle is taxed') accounting for more than 600 million alone\footnote{This high number is driven by a variety of factors, for example automatic number plate recognition cameras that can also automatically check vehicle history, and frequent mandatory checks by car insurers, amongst other factors.}. The median service completes around 44,000 transactions per year. It is worth noting that of these 965 million transactions, approximately 866 million are from services where we had the true figure available, and only 99 million are from our predictive model, meaning that any errors in the predictive model do not affect the overall results too much. It is also worth noting that this figure is broadly inline with the 2017 Performance Platform which showed 1.03 billion transactions (though this platform had a slightly different definition of services, and counted things such as passenger arrivals at the UK border, which we did not consider). 

Figure \ref{fig:fig2}, Panel D shows how the amount of transactions conducted by services break down by topic. The two driving agencies - the Driver and Vehicle Licensing Authority (DVLA) and the Driver and Vehicle Standards Agency (DVSA) -  both under the Department of Transport (DfT), have the highest volumes, likely driven by the many transport-related services the government provides, as well as the considerable variety of regulations around driving especially. The Department for Work and Pensions (DWP) is also notable for having a high volume of transactions. 

Despite these high absolute numbers, it is worth emphasising that they of course represent an under-counting of the true volume of transactions conducted by central government when thinking about service provision as a whole. As we have highlighted above, we do not claim that our dataset is a complete record of all UK government services (though we believe it is as complete as it can be). Furthermore, we address only transactions required to conduct a `new' service with government, but not those that might be required to maintain an existing arrangement. For example, we include approximately 444,000 new Personal Independence Claims per year in our data; but our data do not reflect the almost 1.4 million existing claims, which also require a variety of transactions throughout the year. Finally, we do not consider local government or the wider areas of public service such as the health service. Although we are unable to materially demonstrate this given the fragmented and incomplete nature of government service data writ large, we are confident that a full accounting of all transactions conducted across the UK government would be in the tens of billions at least, considering our data already shows almost one billion transactions. 

\subsection*{The high share of routine tasks in decision-based public services}

\begin{figure}[!htbp]
\centering
\includegraphics[width=\textwidth]{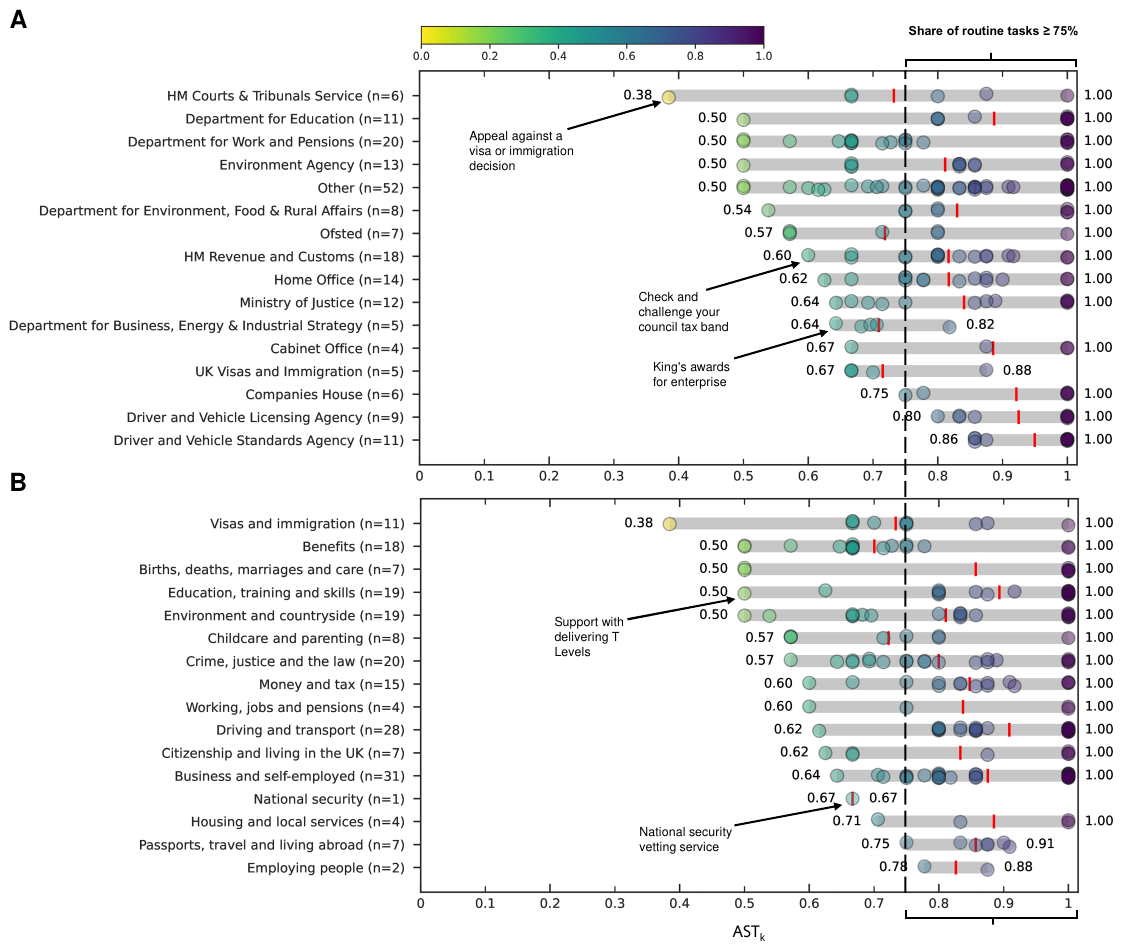}
\caption{\textbf{Measuring the automatability of public services.} Distribution of AST index scores across decision-based services (n = 201) grouped by (\textit{A}) department and (\textit{B}) topic. Services with an AST index score $\geq 0.75$ (equivalent to a share of routine tasks $\geq 0.75$ ) are considered automatable, as indicated by the dotted black line. The red lines show the mean AST index score for each organisation and topic, respectively. Selected services are annotated.}
\label{fig:fig3}
\end{figure}

In this section, we move on to discussing our measurement of the potential automatibility of services. As discussed above, in this analysis, we limit ourselves to 201 decision-based services. Together, these services account for approximately 15\% - or 143 million - of the total 965 million identified service transactions undertaken by government per year. Although a fraction of the total service transactions, our study shows that this 15\% subset of transactions are made up of tasks which are routine and internally time consuming to the public sector \cite{office_for_national_statistics_time_2024}, and therefore can offer a large payoff if automated.

\begin{table}[t!]
    \captionsetup{justification=raggedright, singlelinecheck=false, font=footnotesize}
    \renewcommand{\arraystretch}{1.4} 
    \begin{footnotesize}
    \caption{Distribution of AST index scores}
    \label{table:table2}
    \centering
    \begin{tabularx}{\textwidth}{lcccc}
        \toprule
        \textbf{AST} & \textbf{No. of services} & \textbf{No. of services (\%)} & \textbf{No. of organisations} & \textbf{Total no. of service delivery professionals\textsuperscript{*}} \\
        \midrule
        0 & 0 & 0 & 0 & 0 \\
        $<$ 0.5 & 1 & 0.49 & 1 & 10,430 \\
        $\geq$ 0.5 & 200 & 99.5 & 46 & 76,233 \\
        $\geq$ 0.75 & 149 & 74.13 & 41 & 75,523 \\
        1 & 66 & 32.84 & 30 & 70,133 \\
        \bottomrule
    \end{tabularx}
    \caption*{\textit{Note:} \textsuperscript{*}Service delivery professional numbers are estimates for the number of Administrative Officers and Assistants employed in the UK government Operational Delivery Profession. Data are from 2021 and should be interpreted with caution, and as approximations, as records are missing for key government organisations. Furthermore, not every professional in each organisation is involved exclusively in the provision of decision-based services.}
  \end{footnotesize}
\end{table}

As described, each of the services was reviewed by a team member and tasks were assigned on the basis of the steps that we perceived had to be conducted to complete the task. For the 201 services, the average number of tasks was 7.1 (SD=3.5, min=1, max=230). The most common tasks were `recording, preparing, sorting, classifying and filing information', `verifying the accuracy of information provided', `filing and storing completed documents on computer hard drive or disk', and 'maintain a computer filing system to store, retrieve or update documents', with each assigned to over 100 different services. 

The tasks themselves were then assigned to the meta-task categories to measure the proportion of routine tasks within them. We assume, following the literature, that routine tasks are those that are most amenable to automation (we should note that we have no measurement of the proportion that are \textit{already} automated). The proportion of routine and non-routine tasks is then used to calculate the AST score as described in \hyperref[fig:fig1]{Fig.~\ref*{fig:fig1}\textit{B}}.

\begin{figure}[!htbp]
\centering
\includegraphics[width=\textwidth]{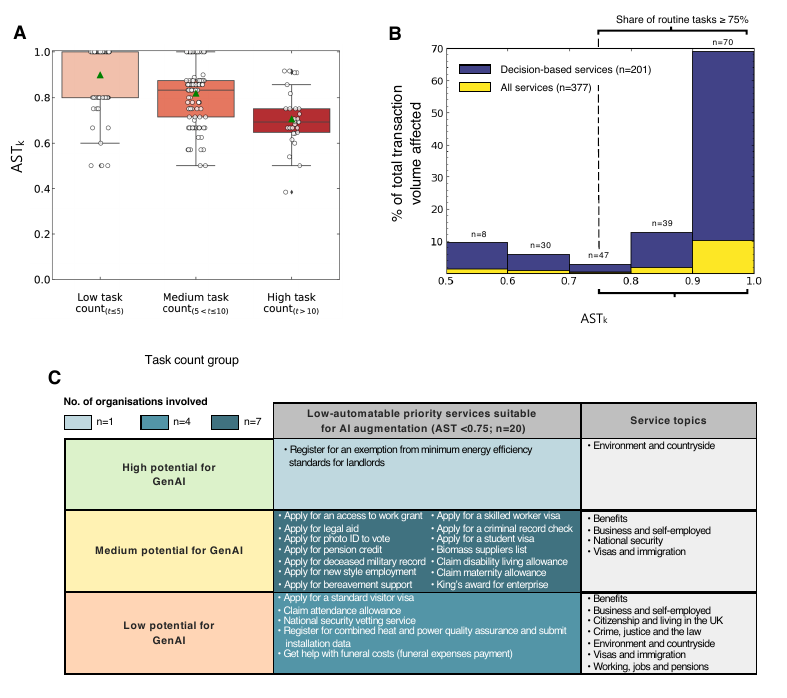}
\caption{\textbf{Predictors of automatability and estimates of potential benefit from adopting generative AI.} (\textit{A}) Differences in the AST distribution based on differences in service task counts. Whiskers are 1.5 times the interquartile range. The solid black line inside each box indicates the median and the green triangles indicate the mean. (\textit{B}) Histogram of AST index scores plotted against transaction volumes shown as a percentage of total volume across only decision-based services ($n=201$) and all services ($n=377$). Numbers above bars show count of services for each category. For illustrative purposes, we only show the range [$0.5,1.0$]. (\textit{C}) Low-automatable decision-based services prioritised by the UK government ($n=20$), defined as having an AST index score $< 0.75$, grouped by potential for benefiting from the adoption of generative AI (Column 1). Column 2 shows the topics associated with the list of services for each segment. Cells are colour coded according to the number of government organisations involved in the delivery of listed services, as per the legend on the top left of the figure.}
\label{fig:fig4}
\end{figure}

The results of this exercise are broken down in  \hyperref[fig:fig3]{Fig.~\ref*{fig:fig3}}. Panel A of this figure shows the distribution of AST scores per department. The graphic highlights some of the least automatable services, such as 'Appeal against a visa or immigration decision', that we might expect to necessarily retain a considerable human component. But it also shows that a considerable amount of services fall above the 75\% threshold that, from previous literature, is considered automatable. It also shows considerable variation per department, with some departments providing highly automatable services according to our measure. For example, all 20 services from the DVLA as well as the DVSA are considered automatable. Figure \ref{fig:fig3}, Panel B shows the same data per topic. Here, the topics of 'driving and transport' as well as 'education, training and skills' appeared to be the most automatable, while the topics of 'benefits', 'childcare and parenting', and 'national security' appeared to be the least automatable. Figure \ref{fig:fig4}, Panel A, meanwhile, looks at the relationship between the volume of tasks and the AST score for each type of service: services with a lower task count are, perhaps unsurprisingly, typically more automatable. 

Table \ref{table:table2} presents a complete breakdown of AST scores, and also shows the number of operational delivery professionals working in departments associated with each of the categories of AST score\cite{cabinet_office_summary_nodate}. This provides a very approximate picture of the current amount of labour required to deliver these services, though we should note this is only a rough guide and not all of the professionals associated with these departments work directly on the services we are studying. We would emphasise that we do not believe the labour of these professionals is directly replacable by AI (very few occupational categories can or should be entirely replaced by technology, especially in public services). However, we do believe this technology could significantly enhance individual and collective departmental productivity by streamlining bureaucratic tasks, freeing up time for public servants to focus on those tasks requiring human judgement, creativity, discretion, and decision-making. 

It is also worth considering these data in terms of the volume of transactions affected. Figure \ref{fig:fig4}. Panel B explores this, showing how the 965 million transactions (represented by the n=377 services) we are considering break down over different proportions of routine tasks. 

We can see that approximately 69\% of the total services transaction volume (n=70 services containing roughly 90 million transactions) fall into the category of between 90 and 100\% fully routine tasks, whilst 84\% of transactions are above the 75\% threshold considered highly automatable. Again, this is significant because it shows the scale of the opportunity: focusing automation efforts here is likely to reap significant dividends. Saving even one minute per transaction in the category of services considered highly automatable on average would save the equivalent of approximately 2 million working hours per year, or around 1,200 years of work, based on the amount of time worked per year in a standard UK full time job\footnote{This is based on a standard working week of 36.4 hours, with an average of 46.5 working weeks in a year when holiday is taken into account. Figures are taken from: \href{https://www.ons.gov.uk/employmentandlabourmarket/peopleinwork/earningsandworkinghours/timeseries/ybuy/lms}{https://www.ons.gov.uk/employmentandlabourmarket/peopleinwork/earningsandworkinghours/timeseries/ybuy/lms}}. 

As we have discussed above, in addition to its high potential to automate routine tasks, the emergence of generative AI also presents a challenge to the notion that non-routine tasks are not susceptible to automation. In order to explore this potential in our data, we conducted a further coding exercise using our PGenAI rubric. Looking in detail at 20 of the 201 decision-based services that had a low percentage of routine tasks within them, we largely follow the methodology set out in \hyperref[sec:pgai]{the Potential for Generative AI section}. Of these 20 services, the majority (14 services, 70\%) were found to have a 'Medium potential' for generative AI integration. These services included `Apply for benefits and visas', as well some services for businesses and self-employed individuals. Only one service was found to have a 'High potential': `Register for an exemption from minimum energy efficiency standards for landlords'. Hence, according to these results, the arrival of generative AI shows even higher potential for automation than might be expected solely by analysing the proportion of routine tasks. 

\subsection*{Conclusion}

In this article we have sought to provide a mapping of the scale of citizen-facing, bureaucratic public service transactions, and to provide a measurement of the potential automatability of these transactions. Through a combination of data sources and empirical modelling, we estimate that central government is providing 377 citizen-facing services, and each year these conduct approximately one billion transactions. Of this one billion transactions, 143 million are complex bureaucratic procedures involving exchanges of data and decision-making, conducted by 201 different services. Our measurement of automatability focused on these services, showing that many of them fell above the threshold typically used in labour economics to identify high automatability. 

Any study that tries to characterise a large proportion of state activity will inevitably contain within it some inaccuracies and approximations. It is therefore worth concluding the article by highlighting some of the limitations of the study, and thus pointing the way for future research. First, there is the question of the extent to which our data captures a full picture of the machinery of government. While we believe our list of government services to be the best possible available, we are conscious that it is likely incomplete. There are likely some services that were not captured by any of the methods we used, especially if they only have a very limited web presence. Furthermore, and perhaps more importantly, we only consider the bureaucratic transactions that initiate access to a service, and do not look at transactions required to keep services `ongoing' or to maintain access to them. And of course we only consider central government, and ignore the many other branches of the state (for example the police, the health service, local government, etc.). For these reasons, the transactions and processes we map are likely only a small fraction of the public sector as a whole. 

Second, there is the extent to which our task-based approach is able to accurately characterise government bureaucratic processes. Our work was largely conducted through desk based research without the collaboration of the service providers themselves. Therefore, although we have made our best effort to map tasks to services based on empirical data, we cannot be sure that this mapping has been done perfectly. Furthermore, all task-based approaches are approximations to an extent, and there are multiple areas where these approximations take place. The ISCO task database we make use of, for example, is international rather than tuned to the specific needs of the UK public sector \cite{lewandowski_technology_nodate}. The extent to which tasks can easily be labelled as `routine' or `non-routine' is also frequently debated, as well as how this definition might differ across contexts and occupations \cite{sebastian_routine_2018}. Furthermore, as we suggest with our work on generative AI, the perception of routine tasks as the only components of occupations - and in our case, services - that are amenable to automation is one that is beginning to shift. It is also worth noting that, in the absence of clear time-use data, we assume that an equal amount of time is spent on each task. We therefore assign an equal weighting to all tasks in the final analysis, which might result in the over- or underestimation of automation potential \cite{dengler_impacts_2018}. For these reasons, further work within government that attempts a more granular task-based mapping would be highly valuable, especially work which would involve direct engagement with service delivery professionals to more robustly capture the nature of their day-to-day work. 

\subsection*{Acknowledgments}

We are grateful for comments, insights and additional data from Matthew Lyon and Clifford Sheppard from the Central Digital and Data Office (CDDO) and the Government Digital Service, respectively. We are also thankful for instructive comments by Amelia Armstrong from CDDO and participants of the DataConnect23 conference. We thank members of The Alan Turing Institute's Public Policy Programme for valuable feedback in an early presentation of this research, as well as the reviewers of IC2S2'2023 and the Workshop on AI Systems for the Public Interest at KI2023. This work was supported by Towards Turing 2.0 under the EPSRC Grant EP/W037211/1 and The Alan Turing Institute.

\subsection*{Author contributions}

V.J.S. collected all secondary data, designed the AST index, performed the analysis and designed and produced the figures and tables. V.J.S. and Y.H. designed the PGenAI rubric. J.B. and J.F. built the model for estimating missing transaction volumes. All authors contributed to the manual assignment of tasks to services. V.J.S, Y.H. and J.B. interpreted the results and drafted the manuscript. All authors approved the final manuscript.

\subsection*{Competing interests}

The authors declare no competing interests.

\newpage

\bibliography{references_3}

\end{document}